\documentclass{elsart}
\usepackage{graphicx}
\journal{Solid State Communications}

\begin{document}

\begin{frontmatter}

\title{Origin of Magnetoresistance Anomalies in Antiferromagnetic
$YBa_{2}Cu_{3}O_{6+x}$}

\author{A.S. Moskvin, Yu.D. Panov}

\address{Department of Theoretical Physics, Ural State University, 620083
Ekaterinburg, Russia}

\begin{abstract}
Specific $d$-wave angular dependence of the in-plane magnetoresistance in
antiferromagnetic tetragonal YBa$_2$Cu$_3$O$_{6+x}$ ($x\,\sim 0.3$) on the
orientation of the external magnetic field within the ($a,b$) plane is assigned
to the effective hole transport through the low-lying excited purely oxygen
doublet O2pe$_u$ state in CuO$_4$ plaquette, rather than the ground
$b_{1g}(d_{x^2-y^2})$ state. The external magnetic field is believed to
determine the orientation of the strong exchange field for the spin-triplet
$b_{1g}e_u:{}^{3}E_u$ state of the hole CuO$_4$ center and owing to the
spin-orbital coupling result in a specific orbital polarization of the $E_u$
doublet. Namely this  gives rise to the spatial d-wave like anisotropy of the
hole transport. The experimental data allow to estimate the parameter of the
effective spin Hamiltonian.
\end{abstract}
\end{frontmatter}

\section{Introduction}

The transport properties of a single hole in a strongly correlated
antiferromagnetically ordered quasi-2D cuprate have been the topic of much
debate, both theoretically and experimentally.  Unusual magnetoresistance
anomalies in the heavily underdoped antiferromagnetic
YBa$_{2}$Cu$_{3}$O$_{6+x}$ ($x=0.30;0.32$) crystals were reported recently by
Y. Ando {\it et al.} \cite{Ando}. The in-plane resistivity $\rho_{ab}$
exhibits unconventional metal-dielectric duality with the high-temperature ($T
> 50\,$K) metal-like behavior in contrast with the low-$T$ insulating one which
is not compatible both with that  for a simple band insulator and for an
Anderson insulator.

The crystals demonstrate an unusual behavior of the in-plane magnetoresistance,
$\Delta \rho_{ab}/\rho_{ab}$ when the magnetic field ${\vec H}$ is applied
along the CuO$_2$ planes. These are a striking d-wave shaped ($\propto
\cos2\phi$) angular dependence, anomalous low-field behavior with saturation
above a well-defined threshold field, and hysteretic effects at low
temperatures. As the temperature decreases below $20-25\,$K, the magnetic field
dependence of $\rho_{ab}$ becomes essentially irreversible, and the system
acquires a memory. The application of the field results in a persistent change
in the resistivity. The authors consider qualitatively these features to be a
manifestation of the 'charge stripe' ferromagnetic structure in this system,
which could be easily  rotated by a rather small external magnetic field. As
the temperature is lowered, it is expected that the stripe dynamics slows down
and the appropriate texture in the CuO$_2$ layers is frozen, forming somewhat
like a cluster spin glass.

The antiferromagnetic domain structure in insulating YBa$_2$Cu$_3$O$_{6+x}$
($x\,<\,0.15$) with 1\% Gd$^{3+}$ as a ESR probe has been recently studied in
detail by A. Janossy {\it et al.} \cite{Janossy1,Janossy2}. According to the
ESR data the easy axes of the antiferromagnetic order are along [100], and
[110] are hard axes in the ($a,b$) plane. In zero magnetic field the single
crystal consists of  the equal amounts of domains oriented along two possible
easy axes. External magnetic field in the ($a,b$) plane induces a spin-flop
reorientation transition, and at $h\,>h_{sf}\,=\,5\,$T ($T\, \sim \,20\,$K)
practically all domains arrange perpendicular to the field. The authors
\cite{Janossy1,Janossy2} consider the two models of domain texture: 1) magnetic
domains separated by charge domain walls within the ($a,b$) plane; and 2)
perfect antiferromagnetic plane domains separated by defective ($a,b$) planes,
and they conclude their ESR data qualitatively evidence in favor of the second
model.

In our opinion, the experimental data \cite{Ando} are unlikely to be consistent
with a  conventional model of the well isolated spin and orbital Zhang-Rice
(ZR) singlet ${}^{1}A_{1g}$ \cite{ZR} believed to be a ground state of the
hole-doped CuO$_4$ center in the CuO$_2$ layers. Here, it should be noted that
when speaking of a Zhang-Rice singlet as being 'well isolated', one implies
that the ${}^{1}A_{1g}$ ground state for the CuO$_4$ plaquette with the two
holes of the $b_{1g}(d_{x^2-y^2})$ symmetry is well separated from any other
excited two-hole states. Indeed, a simple s-like spin and orbital symmetry
implies the tetragonally isotropic 's-wave' transport properties of the well
isolated ZR singlet in the CuO$_2$ layers. The unconventional d-wave
magnetoresistance anisotropy displayed by insulating cuprates, as well as many
other experimental data and theoretical model considerations evidence in favor
of the more complicated structure of the valence multiplet for the hole-doped
CuO$_4$ center rather than simple ZR singlet albeit namely the latter  is a
guideline in the overwhelming majority of current model approaches.

The nature of the valent hole states in doped cuprates is considered as being
of great importance for the high-$T_c$ problem. Having solved the problem we
could justify the choice of the relevant effective Hamiltonian together with
the opportunities of a mapping to the single band $t-J$ or Hubbard model.

Below we show that the model of the valence $^{1}A_{1g}-{}^{1}E_{u}$ multiplet,
developed in
Refs.\cite{Moskvin11,Moskvin12,Moskvin13,Moskvin2,Moskvin31,Moskvin32,Moskvin33},
provides a consistent explanation of the  d-wave magnetoresistance anisotropy
in the CuO$_2$ layers of the doped cuprates believed to be the most important
result of Ref.\cite{Ando}.

\section{The model of the valence
$^{1}A_{1\lowercase{g}}-{}^{1}E_{\lowercase{u}}$ multiplet}

The model implies a quasi-degeneracy in the ground state of the two-hole
CuO$_{4}^{5-}$ center with the two close in energy $^{1}A_{1g}$ and
${}^{1}E_{u}$ terms of $b_{1g}^2$ and $b_{1g}e_u$ configurations, respectively.
In other words, one implies two near equivalent locations for the additional
hole, either to the Cu3dO2p hybrid $b_{1g}(d_{x^2-y^2})$ state to form  ZR
singlet ${}^{1}A_{1g}$, or to purely oxygen nonbonding doublet $e_{u}x,y$ state
with peculiar Cu$^{2+}$-Cu$^{3+}$ valence resonance. The electron density
distribution for two valent hole states $b_{1g}$ and $e_u$, respectively,
together with the quantitative picture of the energy spectrum for valence
multiplet is shown in Fig.1. It should be noted that the symmetry of the
O2pe$_{u} x,y$ states coincides with that of for Cu4p$x,y$ states.

\begin{figure}
  \includegraphics{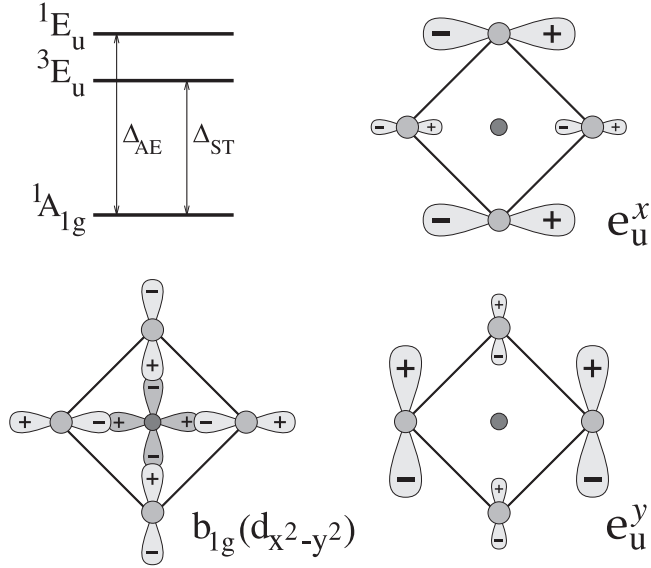}
\caption{\label{fig1}Energy spectrum of the valence
$(b_{1g}^{2}){}^{1}A_{1g}-(b_{1g}e_{u}){}^{1,3}E_{u}$ multiplet and electron
density distribution for valent $b_{1g}$ state and for lower in energy
antibonding $e_u x,y$ combinations of $e_{u}^{\pi} $ and $e_{u}^{\sigma}$
orbitals}
\end{figure}

In a sense, the valence $(b_{1g}^{2}){}^{1}A_{1g}-(b_{1g}e_{u}){}^{1}E_{u}$
multiplet for the hole CuO$_{4}^{5-}$ center implies an unconventional state
with $Cu$ valence resonating between Cu$^{3+}$ and Cu$^{2+}$, or
'ionic-covalent' bonding \cite{Goodenough}. In other words, the CuO$_4$ center
with the valence $(b_{1g}^{2}){}^{1}A_{1g}-(b_{1g}e_{u}){}^{1}E_{u}$ multiplet
represents a specific version of the 'correlation' polaron, introduced by
Goodenough and Zhou \cite{Goodenough}.

The model  is supported both by local-density-functional calculations
\cite{McMahan1}, {\it ab initio} unrestricted Hartree-Fock self-consistent
field MO method (UHF-SCF) for copper-oxygen clusters \cite{Tanaka,Tanaka1}, and
a large variety of experimental data. To the best of our knowledge the one of
the first quantitative conclusions on a competitive role of the hybrid
copper-oxygen $b_{1g}(d_{x^2-y^2})$ orbital and purely oxygen O2p$_{\pi}$
orbitals in the formation of valent states near the Fermi level in the CuO$_2$
planes has been made by A.K. McMahan {\it et al.}  \cite{McMahan1} and J.
Tanaka {\it et al.} \cite{Tanaka}.  Namely these orbitals, as they state,
define the low-energy physics of copper oxides.

One of the most exciting experimental evidences in favor of the model with the
valence $^{1}A_{1g}-{}^{1}E_{u}$ multiplet is associated with the observation
of the midinfrared (MIR) absorption bands which polarization features are
compatible with those for $^{1}A_{1g}-{}^{1}E_{u}$ intra-multiplet dipole
transitions \cite{Moskvin2}. The corresponding transition energies ($\sim
\Delta _{AE}$) observed for various cuprates are of the order of a few tenths
of eV, that yields a typical energy scale for the valence multiplet.

The $e_u$ hole can be coupled with the $b_{1g}$ hole both antiferro- and
ferromagnetically. This  simple consideration indicates clearly a necessity to
incorporate in the valence multiplet both the spin singlet
$(b_{1g}e_{u}){}^{1}E_{u}$ and the spin triplet $(b_{1g}e_{u}){}^{3}E_{u}$,
which energy could be even lower due to ferromagnetic   $b_{1g}-e_{u}$
exchange. Indeed, the low-lying spin triplet state for the two-hole
CuO$_{4}^{5-}$ center was detected by $^{63,65}$Cu NQR in
La$_{2}$Cu$_{0.5}$Li$_{0.5}$O$_{4}$ with a singlet-triplet separation
$\Delta_{ST} = 0.13$ eV \cite{Fisk}. The indirect manifestations of  O2p$\pi$,
or $e_u$ valent states  were detected in the Knight shift measurements by NMR
for 123-YBaCuO system \cite{Yo}. In connection with the  valence
$^{1}A_{1g}-{}^{1,3}E_{u}$ multiplet model for copper oxides one should note
and comment the results of paper by Tjeng et al. \cite{Tjeng}, where the
authors state that they "are able to unravel the different spin states in the
single-particle excitation spectrum of antiferromagnetic  CuO and show that the
top of the valence band is of pure singlet character, which provides strong
support for the existence and stability of Zhang-Rice singlets in high-$T_c$
cuprates". However, in their photoemission studies they made use of the
Cu2p$_{3/2}$($L_{3}$) resonance condition that allows to detect unambiguously
only copper photo-hole states, hence they cannot see the purely oxygen
photo-hole $e_u$ states.

It should be noted that the complicated $^{1}A_{1g}-{}^{1,3}E_{u}$ structure of
the valence multiplet for the two-hole CuO$_{4}^{5-}$ center could be revealed
in the photoemission spectra, all the more that the odd ${}^{1}E_{u}$ terms
play here a principal role, namely these yield a nonzero contribution to the
ARPES for ${\vec k}=0$, or in other words at  $\Gamma$ point. In this
connection one should note the experimental measurements of the photoemission
spectra in Sr$_2$CuO$_2$Cl$_2$ \cite{Wells,Durr} and Ca$_2$CuO$_2$Cl$_2$
\cite{Ronning1,Ronning2}. All these clearly detect a nonzero photocurrent
intensity in the BZ center, thus supporting the $^{1}A_{1g}-{}^{1,3}E_{u}$
structure of the ground state valence multiplet. Overall, the model of extended
$^{1}A_{1g}-{}^{1,3}E_{u}$  valence multiplet allows to  explain consistently
many puzzling properties both of insulating and superconducting cuprates:
midinfrared   absorption bands \cite{Moskvin2}, (pseudo)Jahn-Teller (JT) effect
and related phenomena \cite{Moskvin31,Moskvin32,Moskvin33},  spin properties
\cite{Moskvin41,Moskvin42}.

\section{Anomalous magnetotransport in
$\lowercase{b}_{1\lowercase{g}}\lowercase{e}_{\lowercase{u}}:{}^{3}E_{\lowercase{u}}$
hole state}

The  pseudo-JT polaronic nature of the spin-singlet $^{1}A_{1g}-{}^{1}E_{u}$
ground state \cite{Moskvin31,Moskvin32,Moskvin33} favors their localization. In
addition, one should account for the antiferromagnetic background which leads
to the crucial enhancement of the effective mass for the moving spin singlets.
So, a spin-singlet small pseudo-JT polaron as a hole ground state is likely to
be immobile. In such a situation the most effective  channel for the hole
transport could be related to the low-lying excited spin-triplet
$b_{1g}e_{u}:{}^{3}E_u$ term. This gives rise to a thermo-activated hole
conductivity actually observed in most of slightly doped cuprates.

In order to obtain  magnetoresistivity effect we'll consider spin and
spin-orbital interactions for the spin-triplet $b_{1g}e_{u}:{}^{3}E_u$ state.
The $e_u$ hole is strongly exchange-coupled both with the $b_{1g}$ hole on the
same CuO$_4$ center and with the nearest neighboring CuO$_4$ centers. The spin
state of the isolated CuO$_4$ center with the $b_{1g}e_u$ hole configuration is
described by two spin operators
\begin{equation}
  \vec{S}=\vec{s}_{b_{1g}}+\vec{s}_{e_{u}}\,,\quad
  \vec{V}=\vec{s}_{b_{1g}}-\vec{s}_{e_{u}}
\end{equation}
(${\vec S}^{2}+{\vec V}^{2}=\, 3/2,\,\, ({\vec S}\cdot {\vec V})=0$), and
corresponding order parameters \cite{Moskvin41,Moskvin42}. For the description
of the orbital $E_u$ doublet one could use  the pseudospin $s=1/2$ formalism
with the Pauli matrices $\sigma _{x,y,z}$ having simple transformation
properties within the $E_{u}x,y$ doublet:\footnote{We make use of standard
notations for the $D_{4h}$ point group representations.}
$$
  \sigma_{x} = \pmatrix{ 0 &  1 \cr 1 & 0 }\propto b_{2g}\,,\;\;
  \sigma_{y} = \pmatrix{ 0 & -i \cr i & 0 }\propto a_{2g}\,,\;\;
  \sigma_{z} = \pmatrix{ 1 & 0 \cr 0 & -1 }\propto b_{1g}\,.
$$
Then the effective spin-Hamiltonian for the spin-triplet ${}^{3}E_u$  state
of the hole center can be represented as follows
\begin{equation}
  \hat H_S = \lambda S_z \sigma _{y} + D S_{z}^{2} +
    a(S_{x}^{2}-S_{y}^{2})\sigma_{z} + b \widetilde{S_{x}S_{y}}\sigma _{x}
    - \mu_{B}{\vec h}\hat{g}_{S}{\vec S} - 2\mu _{B}{\vec H}_{ex}{\vec S} , \;\;
 \label{H_S}
\end{equation}
where $\widetilde{S_{x}S_{y}}=1/2(S_{x}S_{y}+S_{y}S_{x})$, $\lambda$ is an
effective spin-orbital coupling parameter for the $^{3}E_u$ term, $D,a,b$ the
spin anisotropy constants, $\hat{g}_{S}$ the effective $g$-tensor,
$\vec{H}_{ex}$ the internal effective spin exchange field, $\vec{h}$ the
external magnetic field. Hereafter, we will assume for simplicity the ideal
planar CuO$_2$ layers, isotropic spin $g$-factor  and the planar direction of
magnetic fields. In the frames of the strong molecular field approximation the
spin operators can be replaced by appropriate averages
$$
  \left\langle S_{x}^{2}-S_{y}^{2} \right\rangle = \cos2\Phi \,,\quad
  \left\langle \widetilde{S_{x}S_{y}} \right\rangle = \sin2\Phi \,,
$$
and the spin-Hamiltonian (\ref{H_S}) transforms into an effective Hamiltonian
of the spin-induced low-symmetry crystalline field
\begin{equation}
  \hat H_S = a \cos2\Phi \,\sigma_z + b \sin2\Phi \, \sigma_x \,,
  \label{Hs}
\end{equation}
where $\Phi$ is the azimuthal  angle of the  orientation for the total magnetic
field ${\vec H}={\vec H}_{ex}({\vec h})+{\vec h}$. Eigenvectors and eigenvalues
for such a simple Hamiltonian are
\begin{eqnarray}
  &&
  \Psi _{+}=\cos\alpha \, |x\rangle + \sin\alpha  \,|y\rangle \,,\quad
  \Psi _{-}=\sin\alpha \, |x\rangle - \cos\alpha \, |y\rangle
  \\
  &&
  \tan2\alpha=\frac{b}{a}\tan2\Phi \,,\quad
  E_{\pm}=\pm \Delta (\Phi) \,,\quad
  \Delta(\Phi)=[\,b^2+(a^2-b^2)\cos^{2}2\Phi]^{\frac{1}{2}} ,\nonumber
  \label{eigen}
\end{eqnarray}
respectively, where $|x,y\rangle \equiv |E_{u}x,y\rangle$. The quantum-mechanic
and thermodynamic  averages for $\sigma _{z,x}$, describing the orbital
polarization (quadrupole ordering) effect, are
\begin{eqnarray}
  &&
  \langle \sigma_{z}\rangle_{\pm} =
  \langle \Psi_{\pm}|\sigma_{z}|\Psi_{\pm} \rangle = \pm \cos2\alpha \,,
  \nonumber\\
  &&
  \langle\langle \sigma_{z}\rangle\rangle =
  \rho_{+}\langle\sigma_{z}\rangle_{+} +
  \rho_{-}\langle\sigma_{z}\rangle_{-} =
  -\frac{a\cos2\Phi}{\Delta(\Phi)} \tanh\beta \Delta(\Phi) \,,
  \\
  &&
  \langle\sigma_{x}\rangle_{\pm} = \pm \sin2\alpha, \quad
  \langle\langle\sigma_{x}\rangle\rangle =
  -\frac{b\sin2\Phi}{\Delta(\Phi)} \tanh\beta \Delta(\Phi) \,,
  \nonumber
\end{eqnarray}
respectively. Here, $\rho _{\pm}$ is the statistical weight of the $\Psi_{\pm}$
states, $\beta=1/kT$.  One should note a specific $d$-wave ($d_{x^2-y^2}$ and
$d_{xy}$) angular $\Phi$-dependence for the thermodynamic averages $\langle
\langle \sigma _{z}\rangle \rangle$ and $\langle \langle \sigma _{x}\rangle
\rangle$, respectively (see Fig.2).

\begin{figure}
  \includegraphics{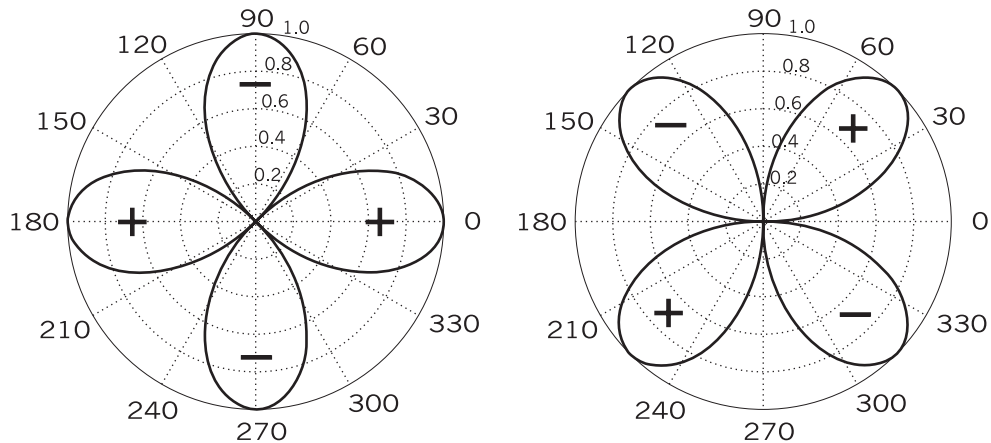}
  \caption{\label{fig2}Polar plot for the high-temperature angular dependence of
the spin-induced quadrupole polarization of the $e_u$ orbitals in CuO$_4$
centers: left - $\langle \langle \sigma _{z}\rangle \rangle $; right - $\langle
\langle \sigma _{x}\rangle \rangle $}
\end{figure}

So, in frames of our approximation, the orbital state of the spin triplet
${}^{3}E_u$ is easily governed by the external magnetic field,  thus  providing
an effective ``magneto-orbital'' transformation. Obviously, this spin-induced
orbital polarization has considerable magnitude only at rather low
temperatures. In the high-temperature limit ($|a|,\,|b|\,\ll \,kT$):
\begin{equation}
  \langle \langle \sigma _{z}\rangle \rangle  \approx
  \,-\frac{a}{kT}\, \cos2\Phi \;.
  \label{6}
\end{equation}
It should be noted that Hamiltonian (\ref{Hs}) possess an overall tetragonal
symmetry, however, this could describe the effects of spontaneous breaking of
this symmetry separately in spin and orbital subspaces, or the effects of the
spin-induced breaking of tetragonal symmetry in the space of orbital $E_u \pm$
states.

Assuming the only nonzero (Cu3d-O2p)$_{\sigma}$-bonding contribution to the
$e_u$ hole transfer between the $E_u $ states on  the nearest neighboring
CuO$_4$ plaquettes, we can represent the matrix of the $e_u -e_u$ hole transfer
integral ${\hat t}(e_u - e_{u})$ on the $|x,y\rangle$ basis set as follows
$$
  {\hat t}_{a}(e_u - e_{u}) = t_{\sigma}\pmatrix{ 1 &  0 \cr 0 & 0 } ,\quad
  {\hat t}_{b}(e_u - e_{u}) = t_{\sigma}\pmatrix{ 0 &  0 \cr 0 & 1 }
$$
for the transfer in  the [100] and [010] directions, respectively. Here,
$t_{\sigma}=t_{a}(e_{u}x - e_{u}x)=t_{b}(e_{u}y - e_{u}y)$. In frames of the
tetragonal anzatz the $e_u$ hole energy spectrum in the tight binding
approximation with the $nn$ transfer consists of two bands $E^{x,y}$ formed by
the transfer $e_u x-e_u x$ and $e_u y-e_u y$, respectively
\begin{equation}
  E^{x}_{{\bf k}}=2t_{\sigma}\cos(k_{x}a) \;, \quad
  E^{y}_{{\bf k}}=2t_{\sigma}\cos(k_{y}a) \;.
\end{equation}
These 1D bands are characterized by maximally anisotropic effective mass, and
describe the 1D hole motion in $a$- and $b$- directions, respectively.
Nevertheless, with account for rigorous $x-y$ ($a-b$) symmetry it is easy to
see that the transport properties  appear to be tetragonally isotropic. The
spin-induced low-symmetry crystalline field (\ref{Hs}) results in a mixing of
the two bands. Indeed, instead of simple expressions for the bare  hole
transfer integrals  we obtain matrices for the renormalized hole transfer
integrals on the $|\pm\rangle$ basis set
\begin{eqnarray}
  {\hat t}_{a}(e_u - e_{u})&=&t_{\sigma}\pmatrix{ \cos^{2}\alpha  &
  \frac{1}{2}\sin2\alpha \cr \frac{1}{2}\sin2\alpha  & \sin^{2}\alpha \cr},
  \nonumber\\
  {\hat t}_{b}(e_u - e_{u})&=&t_{\sigma}\pmatrix{ \sin^{2}\alpha  &
  -\frac{1}{2}\sin2\alpha  \cr -\frac{1}{2}\sin2\alpha  & \cos^{2}\alpha \cr},\,
\end{eqnarray}
where we have simply made use of Exps.(\ref{eigen}) for the $\Psi _{\pm}$
functions.

Thus, after simple algebra we can obtain for the renormalized bands
\begin{eqnarray}
  &&
  E^{\pm}_{{\vec  k}} = t_{+}({\vec k}) \mp
    \left[ \, t_{-}^{2}({\vec k}) + \Delta^{2}(\Phi) +
    2\Delta (\Phi)\,t_{-}({\vec k})\cos2\alpha \, \right]^{\frac{1}{2}},
  \nonumber\\
  &&
  t_{\pm}({\vec k}) = t_{\sigma}\left(\cos(k_{x}a) \pm \cos(k_{y}a)\right).
\end{eqnarray}

The inverse in-plane effective-mass tensors for the two bands can be written as
follows
\begin{equation}
  \widehat{\left(\frac{1}{{\bf m}^{*}}\right)}_{\pm} =
  \frac{1}{m_{0}^{*}}
    \left[\, {\bf 1}\, \pm \, \cos2\alpha \pmatrix{ 1 &  0 \cr 0 & -1 } \right],
\end{equation}
where $\frac{1}{ m_{0}^{*}}=-\frac{4a^{2}}{\hbar ^{2}}t_{\sigma}$.

Thus, we obtain the spin-induced shift/splitting  of the  bare bands, which
results in the breaking  of the tetragonal $x-y$ symmetry with appearance of
the spin-dependent anisotropy in  effective mass.

Taking into account  the proportionality relation between the single-band
conductivity and the inverse  effective-mass tensor, one might obtain a
surprisingly simple expression for the in-plane magnetoresistance
\begin{equation}
  \frac{\delta \rho_{a,b}}{\rho_{a,b}} =
  \langle \cos2\alpha \rangle =
  \mp  \langle\langle \sigma_{z} \rangle\rangle ,
  \label{mr}
\end{equation}
where the upper (lower) sign corresponds to the [100] ([010]) directions,
respectively. In other words, the in-plane magnetoresistance appears to be
straightforwardly linked to the spin-induced orbital (quadrupole)  polarization
of the $e_u$ states in the CuO$_4$ centers. Rigorously speaking, the expression
(\ref{mr}) is valid in the high-temperature region $\Delta (\Phi) \ll kT$, when
this easily (see (\ref{6})) reduces to
\begin{equation}
  \frac{\delta\rho_{a,b}}{\rho_{a,b}} \, \approx \, \mp \frac{a}{kT} \cos2\phi,
  \label{mr1}
\end{equation}
where, in contrast to (\ref{6}), $\phi$ is an in-plane azimuthal angle for the
external field ${\vec h}$, and we take into account  the near orthogonality of
${\vec H}$ and ${\vec h}$. This extremely simple expression describes all the
essential features of the magnetoresistance anisotropy in the CuO$_2$ layers
and represents a main result of our model theory. Indeed, when comparing with
experimental data \cite{Ando} for $YBa_2Cu_3O_{6+x}$ ($x\,\sim 0.3$) one might
see that the experimental angular dependence of the magnetoresistance for a
rather strong external field exceeding the  $ h_{flop}\approx 5\, T$  obeys the
($\phi , T$) dependence (\ref{mr1}) rather well given the relatively small
albeit reasonable value of the parameter $a$ of the spin anisotropy: $a\,
\approx \,+\, 0.1 K$.

Observed $T$-dependent deviations from a simple $\propto \cos2\phi $ law can be
related to the JT effect within the $E_u$ doublet
\cite{Moskvin31,Moskvin32,Moskvin33} which results in strong spin-vibronic
effects  providing  the spin-induced distortions of the CuO$_4$ center.
Moreover, the JT effect results in a two-well adiabatic potential with the
stabilization of the hybrid structural-orbital modes of the $b_{1g}$, or
$b_{2g}$ symmetry, providing the $\propto \cos 2\phi$, or $\sin 2\phi$
dependence of the magnetoresistance, respectively. These problems will be
considered in details elsewhere.

We did not address important issues related to the origin of magnetic
anisotropy and AF domain textures in doped insulating cuprates. In our opinion,
the decisive role here is played by the doping-induced nucleation of the
domains of a novel phase, which percolation at $x\geq 0.4$ results in
superconductivity \cite{Furrer}. The effective nucleation of these domains in
the CuO$_2$ layers is promoted by a strong in-plane charge inhomogeneity
generated at least by three linearly arranged $nn$ chain oxygen atoms. Namely
this leads to a quasi-1D stripe-like structure of the in-plane domains. The
$[100]$  and $[010]$ oriented  stripe-like domains induce the orthorhombic
crystal structure distortions in the surrounding antiferromagnetic tetragonal
matrix and generate an appropriate in-plane magnetic anisotropy. Thus, the
CuO$_2$ layers of the underdoped tetragonal insulating YBa$_2$Cu$_3$O$_{6+x}$
could be considered as antiferromagnets with the stripe-induced fluctuating
in-plane magnetic anisotropy with the competition of two easy axes $[100]$ and
$[010]$, respectively.

For the well developed phase separation regime with the comparable volume
fractions of both phases the in-plane resistivity could be qualitatively
represented as a sum of three contributions: the semiconducting one, the stripe
contribution, and the contact resistivity of the interface transport. One
should note that the stripe domains in the hole doped cuprates could be
considered as a source of the hole carriers (donors) for the semiconducting
matrix.

The observed effects such as the low-temperature magnetic hysteresis and the
memory could be associated with the quenching of the stripe texture below
$T\approx 20 $ K due to the sharp slowing down of the inter-phase boundary
relaxation revealed by the ${}^{63,65}$Cu NQR measurements \cite{Matsumura}. It
is of interest to note, that the low-temperature ($20\div 25$ K) spin-glass
like transition has been reported for heavily underdoped antiferromagnetic
Y$_{1-y}$Ca$_{y}$Ba$_{2}$Cu$_3$O$_6$ \cite{Niedermayer}.

\section{Conclusion}

In conclusion, we propose a model where specific $d$-wave angular dependence of
the in-plane magnetoresistance in antiferromagnetic tetragonal
YBa$_2$Cu$_3$O$_{6+x}$ ($x\,\sim 0.3$) on the orientation of the external
magnetic field in ($a,b$) plane is assigned to the effective hole transport
through the low-lying excited purely oxygen doublet O2pe$_u$ state, rather than
the ground $b_{1g}(d_{x^2-y^2})$ state. Such an approach implies that the
ground state of the doped CuO$_4$ plaquette in cuprates could have
substantially more complicated singlet-triplet $^{1}A_{1g}-{}^{1,3}E_{u}$
structure instead of the well-isolated simple $^{1}A_{1g}$ ZR singlet. External
magnetic field determines the orientation of the strong exchange field for the
spin-triplet $b_{1g}e_u:{}^{3}E_u$ state of the hole CuO$_4$ center and due to
the spin-orbital coupling results in a specific orbital polarization of the
$E_u$ doublet giving rise to the spatial anisotropy of the hole transport. The
observed $d$-wave  angular dependence of the magnetoresistance on the
orientation of the external magnetic field is associated with the shift and
splitting of   two 1D bands, and the anisotropy of the effective mass which in
turn results from the spin-induced effective low-symmetry crystalline field for
the $e_u$ hole.

\section{Acknowledgements}
The research described in this publication was made possible in part by CRDF
grant No.REC-005. The authors acknowledge a partial support from the Russian
Ministry of Education, grant E00-3.4-280, and Russian Foundation for Basic
Researches, grant 01-02-96404. One of us (A.S.M.) would like to thank for
hospitality Max-Planck Institute f\"{u}r Physik  komplexer Systeme and Institut
f\"{u}r Festk\"{o}rper- und Werkstofforschung Dresden where part of this work
was made.

\end{document}